\documentclass[twocolumn,showpacs,preprintnumbers,prb,aps,superscriptaddress]{revtex4-2}
\usepackage{amsmath}
\usepackage{amssymb}
\usepackage[retain-unity-mantissa=false]{siunitx}
\usepackage{graphicx}
\usepackage{xcolor}
\usepackage[caption=false]{subfig}
\usepackage{placeins}
\usepackage[colorlinks=true, linkcolor=blue, citecolor=blue, filecolor=blue, urlcolor=blue,pdfproducer={.},pdfcreator={.}]{hyperref}
\usepackage{cleveref}
\usepackage{blindtext}
\usepackage{comment}
\usepackage{tikz}
\usetikzlibrary{positioning,arrows.meta,calc}
\usepackage{verbatim}
\usepackage{ulem}
\usepackage{physics}
\usepackage[commandnameprefix=always, commentmarkup=uwave]{changes}
\usepackage{orcidlink}
\usepackage[frozencache]{minted}
\definecolor{bggray}{RGB}{245,245,245}

\usepackage{tcolorbox}
\tcbuselibrary{minted,skins,breakable}
\definecolor{cpptype}{HTML}{B00040}
\newcommand{\ctype}[1]{\textcolor{cpptype}{#1}}

\newtcblisting{codefile}[2][]{
  listing engine=minted,
  minted language=#2,
  minted options={fontsize=\scriptsize, breaklines, linenos=true, framesep=2mm, numbersep = 6pt, escapeinside=||},
  colback=bggray, colframe=gray!30,
  listing only,
  fonttitle=\scriptsize\ttfamily,
  coltitle=black,                
  title=\texttt{#1},
  breakable=false,
  sharp corners,         
  boxsep=2pt,            
  left=14pt, right=2pt, top=2pt, bottom=2pt,  
  before skip=4pt, after skip=4pt,            
  boxrule=0.5pt          
}

\definechangesauthor[name=Markus, color=orange]{M}
\definechangesauthor[name=Tobi, color=blue]{T}
\definechangesauthor[name=Weber, color=purple]{SW}
\definechangesauthor[name=Chris, color=cyan]{C}
\definechangesauthor[name=Chris, color=magenta]{BR}

\makeatletter
\def\@fnsymbol#1{\ensuremath{\ifcase#1\or *\or \ddagger\or
   \mathsection\or \mathparagraph\or \|\or **\or \dagger\dagger
   \or \ddagger\ddagger \else\@ctrerr\fi}}
\makeatother

\begin{document}

    \title{From Code to Figure: A FAIR-Aligned Data Provenance Chain\\for Reproducible Simulation Research in Numerical Physics}
    \author{Markus Uehlein \orcidlink{0000-0002-3193-3749}}
    \email{uehlein@rptu.de}
    \author{Tobias Held \orcidlink{0009-0009-8925-1810}}
    \author{Christopher Seibel \orcidlink{0000-0003-1513-1364}}
    \author{\\ Lukas G. \surname{Jonda} \orcidlink{0009-0005-3694-1755}}
    \author{Baerbel Rethfeld \orcidlink{0009-0008-9921-4127}}    
    \email{rethfeld@rptu.de}
    \author{Sebastian T. Weber \orcidlink{0000-0002-9090-2248}}
    \email{sebastian.weber@rptu.de}
    \affiliation{Department of Physics and State Research Center OPTIMAS, RPTU University Kaiserslautern-Landau, Kaiserslautern, Germany}
    
\begin{abstract}
    Computational physics increasingly depends on large simulation datasets generated by software that remains under active development for many years.
    In such settings, reproducibility requires not only well documented data but also explicit links between code versions, simulation inputs, generated outputs, analysis steps, and published figures.
    Here, we present an integrated workflow for reproducible and FAIR-aligned simulation research in numerical physics.
    We describe how version control, code review, automated testing, structured logging, metadata-rich output, and standardized post-processing can be combined to support traceability from software development to publication.
    The presented concepts demonstrated for one particular simulation framework are broadly applicable to computational physics and other data-intensive areas of scientific computing.
\end{abstract}

\date{\today}

\maketitle

\section{Introduction}
The amount of scientific data increases steadily every year, reflecting the rapid growth of digital research and technology.
At the same time, a significant amount of this data remains difficult to interpret, reuse, or reproduce, leading to repeated effort and avoidable costs in research.
In the European Union alone, ineffective research data management results in additional annual costs of several billion euros, placing a substantial financial burden on research institutions and society~\cite{Roche2015}.

In 2016, Wilkinson \textit{et al.} introduced the FAIR principles for research data management, which state that  data should be \textbf{f}indable, \textbf{a}ccessible, \textbf{i}nteroperable, and \textbf{r}eusable~\cite{Wilkinson2016}.
Findable data is described by rich metadata, assigned a persistent identifier such as a DOI~\cite{ISO26324:2025, DOI}, and indexed in a searchable resource. 
Accessible data can be retrieved via standardized protocols, which may require authentication and authorization.
Interoperable data ``use a formal, accessible, shared, and broadly applicable language for knowledge representation''~\cite{Wilkinson2016}.
Reusable data is released with a clear usage license and conforms to community standards.

For computational research, FAIR data management is fundamentally intertwined with software development.
Simulation output is scientifically meaningful only if the executable state, the runtime configuration, and the processing steps that led to its generation can be reconstructed.
In long-lived software projects, reproducibility therefore depends not only on storing output files but also on preserving an explicit chain from code development to simulation input, generated data products, derived figures, and final publication.

This requirement is particularly important in numerical physics, where simulation frameworks often evolve over many years while continuing to generate new datasets as bases for publications.
This process frequently exceeds the tenure of a single doctoral researcher, so multiple researchers contribute to the same codebase across overlapping and successive periods.
As a consequence, sustainable software practices and structured data management become part of scientific quality assurance, rather than optional engineering overhead~\cite{Merali2010, Wilson2014, Wilson2017, Kelly2009}.

In this manuscript, we present an integrated workflow for traceable and FAIR-aligned simulation research, illustrated using our long-lived \textit{monstr} (\textbf{m}odular \textbf{o}bject-oriented \textbf{n}onequilibrium \textbf{s}pin- and \textbf{t}ime-resolved \textbf{r}elaxation) simulation framework.
The code has been in active development since 2019 and has supported multiple recent studies~\cite{Stiehl2022, Uehlein2022, Seibel2022, Seibel2023, Maigler2024, Held2025a, Held2025b, Held2025c, Uehlein2025, Seibel2025b, Wust2022arxiv, Haeuser2023arxiv, De2025, Wrigge2025, Randolph2026, Roden2026arxiv, Seibel2026arxiv} across partially overlapping research topics.
The distinct contribution of this manuscript is to show how reproducibility in long-lived simulation projects can be supported by connecting software review, automated testing, executable-state capture, metadata-rich data storage, and standardized analysis within an end-to-end provenance chain.
Our aim is therefore not to describe the software internals in detail, but to demonstrate how these elements can be linked in practice from the software state to published figures.
Although implemented in C\texttt{++}, the underlying concepts and methodology can be transferred to other programming languages, tools, or simulation problems.

First, we discuss the development practices that make the executable state traceable and reliable.
We then describe how simulation inputs, outputs, and metadata are stored in a structured way and how this enables consistent links from a simulation run to archived datasets and published figures.

\section{Workflow}

\begin{figure*}[t]
\centering
\includegraphics{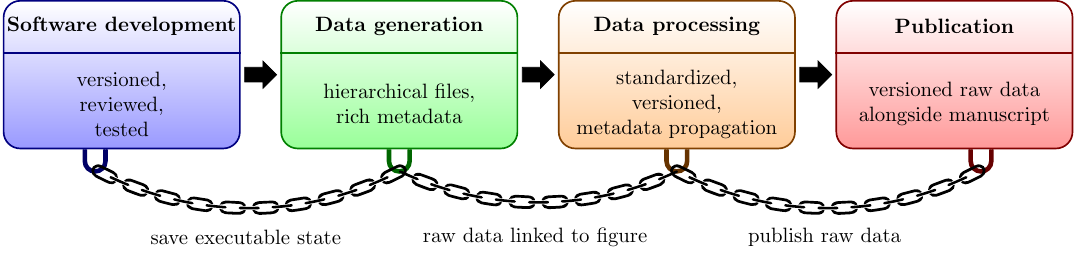}
\caption{
Conceptual overview of the data provenance chain presented in this work.
Version-controlled and reviewed code is built into an executable, which records its exact state in the output file.
The simulation produces a metadata-rich hierarchical file that is consumed by a version-controlled analysis library to generate publication figures.
The metadata of figures contains references to persistent identifiers assigned to the raw data, preserving the provenance chain from code to figure.
To complete this chain, the data should be published alongside the physical results.
}
\label{fig:workflow}
\end{figure*}

We demonstrate the workflow for the \textit{monstr} program, a modular simulation framework to study the dynamics in solids after ultrashort laser excitation.
The framework combines phenomenological temperature-based models~\cite{Mueller2014PRB, Seibel2022, Wust2022arxiv}, kinetic descriptions based on the Boltzmann equation~\cite{Mueller2013PRB, Seibel2023}, optical response functions~\cite{Ndione2022,Seibel2025b,Wrigge2025} and ongoing extensions such as Monte Carlo approaches.
It has been applied to a broad class of materials, ranging from noble metals over ferromagnets to dielectric antiferromagnets and multilayer structures consisting of these materials.

The scientific scope of such a framework creates a challenging reproducibility problem, as the software evolves continuously with the research questions.
Models, material systems, excitation scenarios, and differential-equation solvers must therefore be configurable in a way that is both flexible and consistent.
This gives rise to input-dependent datasets that must be stored in a structured and reusable form.
Moreover, datasets may later be revisited in the context of different physical questions, and a single problem may require combining multiple datasets.
Finally, published figures often depend on additional processing steps beyond the executable and input state that originally produced the underlying data.

Our workflow is designed to address these problems through a provenance chain comprising four connected steps, schematically summarized in \cref{fig:workflow}.
First, the software development process is versioned, reviewed, and automatically and continuously tested (\cref{sec:vc,sec:tests}).
Second, the executable state used for a run is captured together with the simulation input and runtime log, and stored together with the results in a hierarchical file format that supports standardized processing and interpretation (\cref{sec:logging,sec:hdf5}).
Third, data processing and figure generation are performed with a version-controlled analysis library that propagates the relevant metadata into the final documents, linking the extracted data with the underlying raw and metadata~(\cref{sec:figures}).
Fourth, datasets are both stored internally and published with persistent identifiers, so that the full path from software state to publication remains traceable (\cref{sec:publication}).

\subsection{Version control, code review, and executable provenance \label{sec:vc}}

The reproducibility of simulation results requires that the state of the source code used for a calculation can be identified unambiguously, even years after the run.
For this purpose, we use Git as a version control system.
Version control enables simultaneous development of different versions of the code on separate branches, preserves the history of changes, and allows earlier code states to be revisited when simulation results must be checked or reproduced~\cite{Blischak2016, Zolkifli2018}.

For collaborative development, the repository is hosted on a remote GitLab instance.
This provides a central record of code changes and supports branch-based development, issue tracking, and review workflows.
New features are developed in dedicated branches and merged only after a review by another developer.
This dual control principle improves code quality and also creates a documented review history for substantial changes.

To ensure traceability of individual runs, it is not sufficient to record only the repository revision through its commit identifier.
Local, uncommitted source changes present at build time can still affect program behavior.
To address this, our build system automatically records both the current Git commit identifier and the set of local modifications relative to that commit when the executable is built.
In addition, the output metadata includes basic information about the execution environment, such as the CPU model and the MPI (message passing interface) configuration for parallelization.
Together, these records are stored with the simulation output.
Each dataset can therefore be traced back not only to a repository revision but to the exact source state from which the executable was generated.

\subsection{Implementation safeguards and automated quality assurance \label{sec:tests}}

A sustainable and reproducible workflow requires more than version control and executable provenance.
It also depends on implementation safeguards that reduce the chances of silent errors and improve numerical robustness.
In our workflow, these safeguards include compile-time dimensional analysis, numerically stable internal representations, and automated tests.
They are executed automatically through a GitLab continuous integration pipeline whenever changes are pushed to the remote repository.
For the main branch, the same pipeline also rebuilds the documentation generated with \href{https://www.doxygen.nl/}{Doxygen}, keeping the user-facing documentation synchronized with the evolving codebase and making documentation maintenance part of the development workflow.

One common source of implementation errors is the translation of physical equations into code.
Missing constants, inconsistent unit systems, or incorrect algebraic manipulations can alter the numerical result, but they are difficult to detect.
To reduce this risk, we use the \textit{Boost.Units} library~\cite{boost} for dimensional analysis at compile time.
This assigns physical types to quantities, allowing the compiler to reject dimensionally inconsistent expressions.
The following code excerpt illustrates the approach.
\begin{codefile}[types.hpp]{c++}
#ifdef DIMENSIONAL_ANALYSIS
#include "boost/units/systems/si.hpp"
using namespace boost::units;
typedef quantity<si::energy, double> |\ctype{Energy}|;
inline constexpr |\ctype{Energy}| u_energy = 1.0*si::joule;
typedef quantity<si::volume, double> |\ctype{Volume}|;
inline constexpr |\ctype{Volume}| u_volume = 1.0*si::cubic_meters;
#endif

#ifndef DIMENSIONAL_ANALYSIS
typedef double |\ctype{Energy}|;
inline constexpr |\ctype{Energy}| u_energy = 1.0;
typedef double |\ctype{Volume}|;
inline constexpr |\ctype{Volume}| u_volume = 1.0;
#endif

typedef decltype(u_energy/u_volume) |\ctype{EnergyDensity}|;
inline constexpr |\ctype{EnergyDensity}| u_energy_density = u_energy/u_volume;
\end{codefile}
The excerpt defines types and reference units for the physical quantities energy and volume.
When the \texttt{DIMENSIONAL\_ANALYSIS} flag is enabled, these types are aliases for the \textit{Boost.Units} types.
Derived quantities can then be defined directly from these unit-aware types, as illustrated by the energy density (see lines 17 and 18).
The reference units are used to convert the input parameters into the expected internal representation.
In production builds, dimensional analysis is disabled and the corresponding types reduce to \texttt{double} (see lines 10 to 15).
This avoids compatibility issues with external libraries and eliminates runtime cost from unit handling in optimized builds.

In addition to dimensional consistency, numerical stability matters for reproducibility.
Our simulations contain quantities that span many orders of magnitude in SI units, which can amplify round-off errors in double precision.
Therefore, we use atomic units internally~\cite{codata_atomic_units} while retaining SI units at the input and output interfaces.
The atomic unit system is based on the Hartree energy $E_\text{H}$, the reduced Planck constant $\hbar$, the elementary charge~$e$, the electron mass $m_\text{e}$, the Bohr radius $a_0$, and the Boltzmann constant $k_\text{B}$.
For typical solid-state physics applications, this keeps numerical values in a more favorable range without making the workflow inconvenient for users.

For some sensitive quantities, dynamical changes that are orders of magnitude smaller than the quantities themselves can be lost due to floating-point rounding.
This effect can be mitigated by storing a reference value separately from the accumulated change.
In our scientific setting, this is a sensible strategy for tracing the evolution of energy distribution functions \cite{Seibel2023,Seibel2025b}.

Automated unit and integration tests provide a third layer of protection, which we implement using the GoogleTest framework~\cite{gtest}.
Beyond conventional software checks, physical integration tests verify that the implemented models exhibit the expected qualitative and quantitative behavior.
The following example presents such a physically motivated consistency test and highlights the combined use of unit-aware types and atomic units.
\begin{codefile}[test\_internal\_energy.cpp]{c++}
TEST_F(ElectronTest, increaseTemperature){
    |\ctype{Temperature}| T = 300./kelvin_per_au*u_temperature;
    electrons.setTemperature(T);
    |\ctype{EnergyDensity}| u = electrons.getInternalEnergy();
    for (|\ctype{size\_t}| i = 0; i < 10; i++){
        T += 50./kelvin_per_au*u_temperature;
        electrons.setTemperature(T);
        EXPECT_GT(electrons.getInternalEnergy(), u);
    }
}\end{codefile}
This test checks that the electron internal energy increases monotonically with temperature.
Here, the temperature starts at \SI{300}{\kelvin} and is increased in ten steps, and the internal energy is required to exceed its initial value at each step.
Temperature updates (line 6) must remain dimensionally consistent, since incompatible types would be rejected at compile time.
The example also shows the conversion from SI units to atomic units used internally (lines 2 and 6).

\subsection{Validated inputs, runtime execution and structured logging \label{sec:logging}}

Long-lived simulation projects typically expose a large number of configuration parameters.
In our case, models, material systems, solver settings, and excitation conditions are specified in input files using the open YAML format~\cite{yaml}.
This flexibility is useful, but it also creates the risk of inconsistent or physically implausible configurations.
We therefore validate user input before launching a calculation and report errors directly through the console.

To improve consistency across simulations, material systems are defined in a separate, version-controlled YAML database referenced by the main input file.
This ensures that common material parameterizations are reused across runs while preserving the possibility of local overrides in the run-specific configuration.
The resulting separation of reusable material definitions from calculation-specific settings reduces duplication and supports maintainable input management.

We monitor runtime data quality using diagnostics such as particle-number and energy conservation.
Deviations from the expected behavior provide immediate physics-based indicators of numerical or modeling problems.
These diagnostics are stored alongside the primary results so they can be evaluated during post-processing.

To document events during runtime, we supplement input validation with structured runtime logging.
Structured logging provides channels and severity levels such as debug, info, warning, and error.
Specifically, we use the \textit{spdlog} library~\cite{spdlog}, which supports compile-time elimination of disabled log levels
and thus supports both routine monitoring and targeted debugging.
To preserve the execution context of a run, the resulting log is stored together with the associated simulation metadata and output data.

\subsection{Hierarchical storage with HDF5 and NeXus \label{sec:hdf5}}

Once a calculation has been executed, the resulting data must be stored in a way that remains interpretable, processable, and reusable.
Plain text formats such as CSV are convenient for simple one-dimensional data, but they reach their limits when datasets become large, multidimensional, or metadata-rich.
We therefore use the hierarchical HDF5-file format as the underlying storage format~\cite{hdf5, Folk2011}.
HDF5 is an open source binary file format that includes several compression algorithms.

HDF5 provides groups and datasets as the basic building blocks of a hierarchical file.
Groups act as containers, while datasets store the actual data as strings, scalars, vectors, matrices, or higher-dimensional arrays.
This allows us to combine simulation results, coordinate axes, auxiliary quantities, and metadata in a single structured file.

\begin{figure}
    \includegraphics{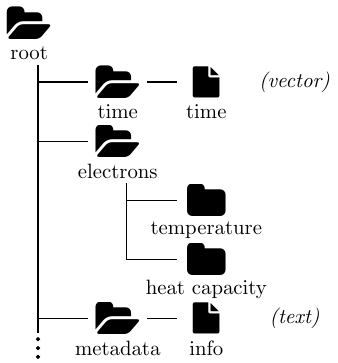}
    \caption{Simplified excerpt of a provenance-rich NeXus output file illustrating its hierarchical organization.}
    \label{fig:structure}
\end{figure}
\begin{figure}
    \includegraphics{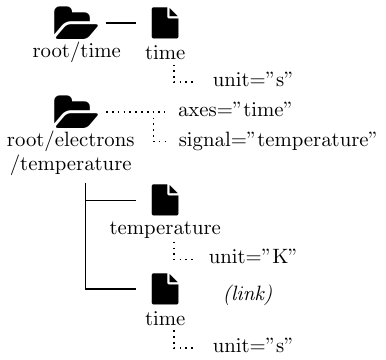}
    \caption{Excerpt from the NeXus structure highlighting the use of attributes and links for documentation, units, and standardized visualization metadata.}
    \label{fig:attributes}
\end{figure}

\Cref{fig:structure} shows a simplified excerpt from a typical HDF5 file produced by our simulations.
A hierarchical file can be understood by analogy with a file system.
Groups and subgroups organize related quantities, while datasets hold the numerical content.
In our workflow, the file contains not only scientific results, but also the metadata needed for traceability.
This includes the Git commit identifier, local code modifications recorded at build time, the YAML input file, the runtime log, and selected execution-environment metadata such as the CPU model and MPI configuration.
Together, this metadata documents the exact source state, input configuration, and execution context of the run.

HDF5 supports links, which are used when the same quantity is required in multiple places.
Instead of storing duplicate copies, it can reference a single dataset from several locations in the file.
This reduces redundancy while maintaining a reader-friendly structure.

HDF5 also allows attributes to be attached to groups and datasets.
We use attributes for documentation and for the standardized description of physical quantities.
We further adhere to the NeXus standard~\cite{Konnecke2015}, developed for neutron, X-ray, and muon science and adapted to solid state physics by FAIRmat~\cite{Scheffler2022}.
NeXus facilitates unified visualization by defining attributes that specify which quantities should be interpreted as signal and axes, and store the corresponding physical units.
\Cref{fig:attributes} shows an excerpt from the output file structure, highlighting the usage of links and additional attributes.

The files can be inspected through H5Web~\cite{h5web} and processed further, for example with the \textit{nexusformat} package in Python~\cite{nexusformat}.
We use the C\texttt{++} HighFive library to write the files from within the simulation software~\cite{highfive}.

By combining an open file format with explicit structure and rich metadata, this output design supports important FAIR objectives.
The data becomes interoperable through HDF5 and NeXus, and more reusable because the context required for interpretation is preserved directly in the file.

\subsection{From simulation output to figures \label{sec:figures}}

In research, published figures are often the most visible endpoint of the workflow.
However, they are only reproducible if they remain linked to both the data products from which they were derived and the analysis steps used to generate them.
For this reason, the output schema of our simulations is standardized across different models as much as possible.
Analysis scripts can therefore operate on the same kind of structured NeXus files regardless of the specific physical model that produced them.

To reduce variability in downstream analysis, we use a separate, version-controlled Python library that provides shared post-processing and plotting routines.
This avoids the use of ad hoc figure-generation scripts and ensures that common derived quantities and visualization steps are implemented consistently across projects.
As a result, the provenance chain does not end with the simulation executable, but extends to the analysis code used to transform simulation output into publication-ready figures.

Because input file, software provenance, runtime log, and execution metadata are recorded inside the NeXus output, post-processing routines can retrieve the relevant identifiers directly from the dataset they consume.
When figures are exported, the commit identifier of the analysis library can, in turn, be embedded in the figure metadata together with a reference to the originating dataset.
Each figure can therefore be linked back to both the simulation output on which it is based and the versioned analysis code used to generate it.
The figure thus becomes part of the same traceable provenance chain rather than a detached presentation artifact.

\subsection{Internal data management and publication \label{sec:publication}}

Traceability also requires a strategy for storing and publishing the generated datasets.
We distinguish between internal project data management during active work and publication-oriented archiving, once results are disseminated.
For internal data handling, Git Large File Storage (LFS) offers a simple and flexible solution~\cite{git-lfs}.
It assigns unique hashes to large files and makes changes to those files traceable within the project workflow.

For publication, data products should be deposited in repositories that provide persistent identifiers and stable access~\cite{repos}.
Suitable services include general-purpose repositories such as \href{https://zenodo.org/}{zenodo.org}, on which we have already published NeXus datasets associated with earlier studies~\cite{Held2025a, Held2025b, Held2025c}.
Domain-specific infrastructures can extend this idea further by connecting active data management and publication workflows, such as the \href{https://nomad-lab.eu/nomad-lab/}{NOMAD} platform for material science~\cite{Scheidgen2023}.

Importantly, publication is not treated as a separate step in which only a reduced subset of processed data is retained.
Instead, the archived dataset remains connected to the same provenance chain established during development and execution.
A persistent identifier then makes this dataset citable and discoverable, while the metadata stored inside the file preserves the information required for interpretation and reuse.

\section{Conclusion}

We have presented a continuous data provenance chain for reproducible and FAIR-aligned simulation research in numerical physics.
Its distinct contribution is to connect software development, simulation execution, structured data storage, and standardized post-processing within an integrated workflow rather than treating these stages as separate tasks.

In our implementation, each simulation output is linked to the exact executable state used to produce it, including the versioned source state, locally modified code recorded at build time, the validated input, and the runtime log.
By storing these elements together with the scientific results in a structured NeXus/HDF5 file, the output remains interpretable and reusable beyond the immediate context of a single calculation.
The same provenance chain can then be continued through standardized post-processing up to the published figure.

The main benefit of this approach is not merely retrospective record-keeping, but a research process in which traceability becomes part of everyday scientific practice.
This is particularly important for long-lived simulation codes, where reproducibility depends on preserving the relation between evolving software, generated datasets, and published results over many years.
Although demonstrated here for a numerical-physics simulation framework, the same principles apply wherever evolving code and complex derived data products shape the research process, including experimental workflows.
In such settings, reproducibility is best supported by maintaining an explicit provenance chain that links code, input, output, analysis, and publication.

\section*{Acknowledgements}
We gratefully acknowledge the support from the Deutsche Forschungsgemeinschaft (DFG, German Research Foundation) through the SFB/TRR-173-268565370 ‘Spin+X’ (Projects A08, B03, and INF).

\section*{Conflict of interest}
The authors have no conflicts of interest to disclose.

\bibliographystyle{IEEEtran}            
\bibliography{bibfile/all.bib, links.bib}

\end{document}